\documentstyle[12pt,titlepage]{article} 
\def\baselinestretch{1.8}
\newcommand{\ba}{\begin{array}}
\newcommand{\ea}{\end{array}}
\newcommand{\bd}{\begin{displaymath}}
\newcommand{\ed}{\end{displaymath}}
\newcommand{\be}{\begin{equation}}
\newcommand{\ee}{\end{equation}}
\newcommand{\bea}{\begin{eqnarray}}
\newcommand{\eea}{\end{eqnarray}}
\newcommand{\no}{\!\!\!\!/}

\def\bra{\langle}
\def\ket{\rangle}

\def\a{\alpha}
\def\b{\beta}
\def\g{\gamma}
\def\d{\delta}
\def\e{\epsilon}
\def\ve{\varepsilon}
\def\l{\lambda}
\def\m{\mu}
\def\n{\nu}
\def\t{\tau}
\def\G{\Gamma}
\def\D{\Delta}
\def\L{\Lambda}
\def\s{\sigma}
\def\p{\pi}

\begin{document}
\begin{titlepage}

\begin{flushright}
\begin{tabular}{l}
MRI-PHY/P970922\\
September, 1997\\
hep-ph/9709392
\end{tabular}
\end{flushright}
\vskip .6cm

\begin{center}
{\Large\bf Filtering Out Signals of Gauge-Mediated Supersymmetry
Breaking: Can We Always Eliminate Conventional 
Supersymmetric Effects ?}\\  
\vskip .6cm

Biswarup Mukhopadhyaya$^1$ and Sourov Roy$^2$\\
{\it Mehta Research Institute,
Chhatnag Road, Jhusi,
Allahabad - 211 019, INDIA} \\
\end{center}
\vskip .5cm

\begin{center}
{\bf ABSTRACT}\\
\end{center}

\noindent We investigate the signal $\g\g + E\no$~ in a high-energy 
linear $e^+e^-$ collider, with a view to differentiating between 
gauge-mediated supersymmetry breaking and the conventional
supersymmetric models. ${\it Prima ~facie}$, there is considerable 
chance of
confusion between the two scenarios if the assumption of gaugino mass
unification is relaxed. We show that the use of polarized electron beams
enables one to distinguish between the two schemes in most cases. There
are some regions in the parameter space where this idea does not work,
and we suggest some additional methods of distinction. We also perform 
an analysis of some signals in the gauge-mediated model, coming from 
the pair production of the second-lightest neutralino.
\hspace*{\fill}
\vskip .2in

\noindent
PACS NOS. : 12.60.Jv, 13.10.+q, 14.80.Ly

\hspace*{\fill}
\vskip .5in

\noindent
$^{1}$E-mail : biswarup@mri.ernet.in \\  
$^{2}$E-mail : ~~~sourov@mri.ernet.in

\end{titlepage}

\textheight=8.9in

\section{Introduction}

The search for supersymmetry (SUSY) \cite{rev} is intimately 
connected with the 
issue of SUSY breaking, since the latter often dictates low-energy
phenomenology, based on which search strategies are devised. In recent
times, a lot of attention has been focused on theories where 
SUSY breaking is conveyed to the ``visible" sector 
through the ordinary standard model (SM) gauge interactions. 
Obviously, one would like to know whether it is possible to 
distinguish between these SUSY breaking schemes and the more
popular ones where gravitational interactions play the decisive role. 
From a phenomenological point of view, this boils down to a distinction
between the experimental signals of the alternative scenarios. 

In models with gauge-mediated SUSY breaking (GMSB) 
\cite{gmsb1,gmsb2,gmsb3,gmsb4,gmsb5} the supersymmetric 
partner of the graviton, i.e., the gravitino ($\tilde G$) is very light 
(practically massless compared to the electroweak scale) and is the 
lightest supersymmetric particle (LSP). The lightest standard model 
superpartner, which is the LSP in the scenario motivated by minimal 
supergravity (SUGRA), now becomes the next to lightest supersymmetric 
particle (NLSP). A natural consequence is that the NLSP can now decay 
into its supersymmetric partner and a gravitino. If the NLSP is a 
neutralino and decays inside the detector, then in collider searches 
one can see signatures of the type $\g\g + E\no$, $\g + E\no$~ etc. 
\cite {gama21,gama22,gama23,gama11,gama12}, 
of which the first one is most promising and well studied. This 
has to be contrasted with the SUGRA-based schemes where the gravitino 
is as massive as the electroweak scale itself, and the LSP is stable.

A considerable amount of effort has been spent to ensure that the $\g\g
+ E\no$ signal for GMSB rises above SM backgrounds \cite {bkgd}. 
It is, however,
equally important to understand whether this signal can ever be mimicked
in the conventional minimal SUSY models (henceforth to be described as
the MSSM case in this paper), and to suggest effective methods of
distinction in such cases. 

In this paper, we confine ourselves to high-energy $e^+e^-$ collision 
experiments. We argue that the $\g\g + E\no$ signals may come from the 
MSSM scenario as well, if we relax the condition of gaugino mass 
unification at an energy scale $\sim 10^{16}$ GeV. In that case, as 
has already been shown in the literature \cite {mele}, 
radiative decays of the
second lightest neutralino $({\tilde \chi^0}_2)$ into a photon and 
the LSP $({\tilde \chi^0}_1)$ is dominant in certain regions of the 
parameter space. Pair-produced ${\tilde \chi^0}_2$'s may then give 
rise precisely to the type of signals mentioned above, sometimes with
comparable strength. 
We suggest two ways of distinguishing the GMSB and MSSM cases, namely,
(i) the use of polarized electron beams and (ii) the distributions in
the total energy of the two emitted photons. And finally, we briefly
study the signals coming from ${\tilde \chi^0}_2$ pairs in the GMSB
scheme. 

The paper is organized as follows. In section 2, we give a brief
description of the minimal model of gauge-mediated supersymmetry
breaking. In section 3, we analyze and compare the two-photon
signal both from GMSB and MSSM. Prospects of signals from 
${\tilde \chi^0}_2$ pairs in GMSB are discussed in section 4.
Section 5 contains our conclusions.

\section{The minimal GMSB model - a brief description}

The minimal model of gauge-mediated supersymmetry breaking \cite {gmsb6} 
requires a
messenger sector consisting of vectorlike quark and lepton superfields
which are coupled to a $SU(3)\times SU(2)\times U(1)$ singlet 
superfield S, through the superpotential 
\bea
W = \l {S} {\overline \Psi} {\Psi}.
\eea
The fields $\Psi$ and $\overline \Psi$ lie in a complete 
$5 + \overline 5$ representation of $SU(5)$ in order to maintain 
gauge coupling unification.\footnote {Actually, up to four generations 
of $5 + \overline 5$ or one $5 + \overline 5$ and one $10 + 
\overline {10}$ are allowed. We consider only one generation here.}  

The scalar (S) and auxiliary ($F_S$) 
components of S acquire vacuum expectation values (VEVs) through 
their interactions with the hidden sector where SUSY is broken 
dynamically. These VEVs induce masses for the messenger fields and 
lift the mass-degeneracy between the messenger fermions and 
sfermions. The breakdown of SUSY is communicated to the visible world 
radiatively via the SM gauge interactions. Thus, the observable 
gauginos and scalars acquire masses at the one-loop and two-loop 
levels, respectively.

The expressions for the masses of gaugino ($M_{\frac 1 2}$) and
scalars ($M_0$) are
\bea
M_{\frac 1 2}(M) = {N_m}~{f_1}\left(\frac\L M \right)~{\frac{{\a_i}(M)}
{4 \pi}}~{\L}, 
\eea
\bea
{M^2_0}(M) = 2~{N_m}~{f_2}\left(\frac\L M \right)~\sum_{i=1}^3
{k_i}{C_i}~{\left(\frac{{\a_i}(M)}{4 \pi}\right)}^{2}~{\L}^2 
\eea
at the scale M, where $M = \l\bra S \ket$ and $\L =\frac {\bra {F_S} 
\ket} {\bra {S} \ket}$.
\noindent $M$ determines the overall scale of the messenger sector; 
$\L$ controls particle-sparticle splitting in that sector as well as
sparticle masses in the observable sector.
\noindent The messenger scale threshold functions are given by
\bea
f_1(x) = \frac{1 + x} {x^2}~{\rm ln}(1 + x) + (x \rightarrow -x)
\eea
\bea
f_2(x) = f_1(x) - \frac{2(1 + x)} {x^2}~\left[{\rm {Li}}_2\left
(\frac{x} {1 + x}\right) - {\frac 1 4}~{\rm {Li}}_2\left
(\frac{2x} {1 + x} \right)\right] + (x \rightarrow -x).
\eea
In Eq.(3), $C_i = 0$ for all gauge singlets and is equal to $\frac 4 3,
\frac 3 4, \left(\frac Y 2\right)^2$ for scalars belonging to the 
fundamental representations of $SU(3)$, $SU(2)$, and $U(1)$, 
respectively. $Y = 2(Q - T_3)$ is the usual weak hypercharge 
and $k_i = 1, 1, \frac 5 3$ for these three groups 
(we do not use grand unification normalization for $\a_1$). 
$N_m$ is the number of messenger generations; in our case $N_m = 1$. 
We have taken into account the contributions from the usual D terms 
and the weak scale threshold corrections while evolving the 
sfermion masses from messenger(M) scale down to the electroweak scale. 
We do not consider the messenger scale threshold corrections 
because they are model dependent. Thus, MSSM phenomenology is 
determined in terms of the parameters $\L,~ x = \frac \L M, 
~{\rm tan}\b$ [the ratio of the two Higgs vacuum expectation 
values(VEVs)] and $\m$ (the Higgsino mass parameter).

\section{Analysis of the two-photon signals}

The two-photon signal may come from a GMSB scenario where  
the lightest neutralino ($\tilde \chi^0_1$) is the NLSP. 
The production of a pair of such NLSPs, followed by each decaying 
to a photon and a gravitino, leads to the signal $$e^+e^- 
\longrightarrow {\tilde \chi^0_1} {\tilde \chi^0_1} 
\longrightarrow \g\g + E\no.$$ Detailed calculations on this signal 
have been performed in many recent works in the contexts 
of the current hadronic \cite {hadron} and $e^+e^-$ colliders 
\cite {ecol} as well as the 
proposed Next Linear Collider (NLC) \cite {nlc}. By far the strongest 
limits on
sparticle masses set so far from such signals are those imposed on the
lightest chargino and neutralino by the D{$\emptyset$} group 
\cite {d0grp}: $m_{\tilde \chi^{\pm}} > 150~GeV$, $m_{\tilde \chi^0_1} 
> 75~GeV$.
We begin by reporting a repeat calculation for an $e^+e^-$ collider 
at $\sqrt s$ = 500 GeV. The SM backgrounds have been removed here 
by using the following set of cuts:
\begin{enumerate}
\item $|M_{inv} - M_Z| \ge 20~GeV$
\item ${(p_T)}_{\g} \ge 20~GeV$
\item $40^0 < \theta_\g < 140^0$ 
\end{enumerate}
where we define $M_{inv}$ as:
\bea
M_{inv}^2 = (P_{e^-} + P_{e^+} - P_{\g_1} - P_{\g_2})^2
\eea
The invariant mass cut eliminates the backgrounds from $e^+e^-
\longrightarrow Z\g\g$, whereas the $p_T$-cuts take care of events such
as $e^+e^- \longrightarrow \n \overline \n \g\g$ through t-channel 
W exchange. Finally, selecting highly central events by imposing a polar
angle cut on each observed photon ensures that the photon is coming 
from the decay of a heavy particle.  
GMSB signals are found to survive these cuts to a large extent.  

In Fig. 1, we plot the cross section of $\g\g$ events as a function of
$\L$ for a center-of-mass energy 500 GeV, for unpolarized,
left-polarized, and right-polarized electron beams, the positrons being 
unpolarized in each case. Only representative combinations of $\m$ and
$\rm {tan} \b$ is chosen for these graphs. As is clear from the previous
sections, the masses of the superparticles increase with $\L$; therefore
a lower limit on $\L$ can be inferred from the mass bounds on the
charginos and neutralinos. Our numerical study extends from this lower
limit to the kinematic limit of $\tilde \chi^0_1$-pair production. Also,
we have used $x = \frac 1 2$ throughout this study.

The plot clearly shows that the cross section
for the right-polarized electron beam is greater than that of 
unpolarized and left-polarized beam. This is expected because, in this 
scenario, the right selectron is much lighter than the left selectron, 
and, as a result, the t-channel contribution for neutralino pair 
production is larger for right than for left-polarized electron beams. 
It should be remembered that at high energies ($\sim 500~GeV$) the 
t-channel contribution dominates over the s-channel one. Moreover, 
the NLSP is dominated by the Bino-component over most of the parameter 
space.

It has been shown \cite {zbr} that the channel ${\tilde \chi^0}_1
\longrightarrow Z\tilde G$ can provide a substantial contribution 
(up to about ${23{\%}}$) to the decay of $\tilde \chi^0_1$. We
have taken this into account in our present calculation by 
multiplying the results with the appropriate branching fractions.

On the other hand, the same two-photon signal can come from the MSSM
scenario when pair-produced $\tilde \chi^0_2$'s decay radiatively into a
photon and an LSP (which is $\tilde \chi^0_1$). The branching ratio (BR)
of this decay can be as large as of the order of 0.9 if one assumes 
that $M_2$ and $M_1$, the $SU(2)$ and $U(1)$ gaugino masses, 
respectively, are free parameters \cite {mele}. This is tantamount 
to at least a 
partial relaxation of gaugino mass unification at a high-energy scale, 
but is nonetheless a feasible scenario in a model-independent analysis.
It has been shown that in order to have a large BR of this radiative 
decay, one needs in general $\m < 0$, $\rm {tan} \b \geq 1$. Further, 
regions where $M_1 \simeq M_2$ and $\rm {tan} \b$ is not much larger 
than 1 are favored.  Whenever $\tilde \chi^0_1$ $(\tilde \chi^0_2)$ 
is gaugino dominated and $\tilde \chi^0_2$ $(\tilde \chi^0_1)$ is 
mainly a Higgsino, the tree-level decay widths of $\tilde \chi^0_2$ get 
suppressed and the radiative decay $\tilde \chi^0_2 \longrightarrow 
\tilde \chi^0_1 \g$ is enhanced. This is termed as dynamical 
enhancement and is discussed in great detail in Ref.\cite {mele}. 
The other 
source of large BR comes into play when $(m_{\tilde \chi^0_2} - 
m_{\tilde \chi^0_1}) \sim 10~ GeV$ and $m_{{\tilde \chi^0_2},
{\tilde \chi^0_1}} \approx M_Z$, causing a strong phase-space 
suppression of the three-body decays. This is called kinematical 
enhancement. 

Taking into account all these effects, we see that the cross section for 
the $\g\g + E\no$ signal can be comparable to that coming from the GMSB
scenario over a large region of the MSSM parameter space. Tables 1 and 2
show these cross sections for unpolarized as well as left- and right-
polarized electron beam. In these tables, we have chosen some
representative points in the ($M_1, M_2$) parameter space where the
branching ratio of the radiative decay $\tilde \chi^0_2 \longrightarrow
\tilde \chi^0_1 \g$ is large and the $\g\g + E\no$ cross section is 
high enough to compare with the GMSB signals shown in Fig. 1. This 
feature is reflected in the neighborhoods of all of these 
representative points, although the functional dependence on $M_1$ 
and $M_2$ is rather complicated.  
The same set of cuts as those applied in Fig. 1 are present here also.
We have taken all the slepton
masses to be degenerate and equal to 120 GeV (Table 1) and 1 TeV
(Table 2). Here we have shown results that are consistent with the
limits on various superparticle masses from the LEP experiments upto
$\sqrt s$ = 172 GeV \cite {lep}. It should be borne in mind, 
however, that such
limits are based on the assumption of gaugino mass unification and are
liable to relaxation once that assumption is withdrawn.

As is evident from these tables, the unpolarized MSSM signal can 
often pass off as the signal of GMSB. This is especially true
in regions close to the kinematic limit of $\tilde \chi^0_1$-pair
production in GMSB or in general in cases where the predicted GMSB
signal rate is on the lower side. As a first solution to the problem, 
we study the signals in both scenarios with polarized electron beams. 
We have already seen that the GMSB signals are jacked up by the 
choice of right polarisation. In parallel, the MSSM signals in a large 
number of cases are larger with left-handed electron. This happens 
whenever $\tilde \chi^0_2$ is dominated by Wino, i.e., the $SU(2)$ 
gaugino component. The number of cases with the reverse effect 
($\tilde \chi^0_2$ is dominated by the Bino component) is relatively 
fewer but still nonvanishing. There are some other regions in the
parameter space where this statement is no longer true, i.e., $\tilde
\chi^0_2$ is dominated by the Wino component but still $A^{\g\g}$
(see, Eq. (7) below) is negative.  This happens because in those cases 
the s-channel contribution becomes important (sleptons masses being 
very high), and the cross sections are crucially controlled by s-t and 
s-u interference, where the signs of the mixing factors play a
significant role.

Thus, if one defines $\s^{\g\g}_{L(R)}$ as the cross section 
for $e^+e^- \longrightarrow \g\g + E\no$ with a left(right)-polarized 
electron, then, if an asymmetry parameter is defined as \cite {gama23} 
\bea
A^{\g\g} = \frac {\s^{\g\g}_L - \s^{\g\g}_R}{\s^{\g\g}_L + \s^{\g\g}_R}
~,
\eea    
the magnitude as well as the sign of $A^{\g\g}$ is a clear measure 
of the sensitivity of the signal to electron polarization. 
This asymmetry is also tabulated in Tables 1 and 2. 
The corresponding asymmetry in GMSB is plotted in Fig. 2, 
which is negative everywhere. So, if $A^{\g\g}$ is positive then it 
comes clearly from MSSM scenario. On the other hand, if $A^{\g\g}$ 
is negative then it is not possible to tell whether the signal comes 
from GMSB or from MSSM.

We, therefore, are led to the conclusion that polarized beams do not 
solve the problem completely, and we are forced to devise some
other method of separating the two scenarios. One rather elegant way is 
to see the combined energy distributions 
$\frac {d\s} {d(E_{\g_1} + E_{\g_2})}$ 
of the two emitted photons in
those particular cases where $A^{\g\g}$ is negative. (Such distributions
have earlier been used \cite {toten} to eliminate SM backgrounds.)
These energy 
distributions are plotted in Fig. 3 for unpolarized electron beam. 
In the case of GMSB, this energy distribution 
is peaked at a point 
$\sim$ 250 GeV ($\frac {\sqrt s} 2$) and wider in comparison to the 
MSSM case, where 
it is sharply peaked at a point much lower than 250 GeV. These 
behaviors can be explained in the following manner. In the GMSB case, 
$\tilde \chi^0_1$ decays into two massless particles. So, the sum of the
two photon energies must be peaked at about a point $\frac {\sqrt s} 2$
which is the missing energy carried by the two $\tilde G$'s. In the case
of MSSM, $\tilde \chi^0_2$ decays into a heavy $\tilde \chi^0_1$ 
(whose mass is close to that of $\tilde \chi^0_2$) and a soft photon. 
The combined photon energy in this case will be peaked at a point much
lower than 250 GeV. The distributions in the two cases have been plotted 
for comparable values of the total cross section. The spread in the 
energy distribution mainly depends on the product of the velocity of 
the decaying particle and the energy of the photon, both measured in the
laboratory frame.
This feature is reflected in Fig. 3. Thus by looking at the 
combined energy distributions of the two photons one can readily 
tell whether the signal comes from GMSB or from the 
conventional SUSY scenario.

\section{Pair-production of $\tilde \chi^0_2$}

In the last section, we have seen that the $\g\g$ + $E\no$~signal 
can as well come from MSSM, and that careful analysis is required 
if one  really wants to distinguish between the two pictures 
through the observation of such events. On the other hand, one 
can study other possible signals of GMSB and try to analyse
whether there are sources which can mimic these. One such signal
is $e^+e^- \longrightarrow \tilde e_R {\tilde e^*}_R \longrightarrow 
l^+l^-\g\g + E\no$~ via a $\tilde \chi^0_1$ which is the NLSP. This 
process has already been studied in several works \cite {slepton}. 
The same signal 
can also come from MSSM through the cascade decays of a pair of 
sleptons. It can be large enough in the regions of the parameter space
where radiative decay of $\tilde \chi^0_2$ is dominant. In GMSB,
the contribution comes mainly from a right-polarized electron
beam via the production of right sleptons, giving rise to a 
negative asymmetry similar to the one defined in the last section.
In the MSSM scenario too, the contributions from right 
sleptons can be greater than those from left sleptons when 
$\tilde \chi^0_2$ is dominated by the Bino component and 
$\tilde \chi^0_1$ is mainly a Higgsino. Here also the asymmetry 
parameter will be negative. Thus there is a possibility that
the above signal also can fake GMSB.

In view of the above, it is desirable to explore further tests of 
GMSB which are unlikely to be faked by an MSSM scenario.
Here we give a brief account of some such signals arising  
from pair-produced $\tilde \chi^0_2$ in an $e^+e^-$ collider 
with $\sqrt s$ = 500 GeV. We list below the processes under study, 
which are governed by the decays of $\tilde \chi^0_2$ \cite {bartl}, 
cascading
into a $\tilde \chi^0_1$ which subsequently goes to a  photon 
and a gravitino:
\begin{enumerate}
\item $e^+e^- \longrightarrow \tilde \chi^0_2 \tilde \chi^0_2
\longrightarrow l^+l^-{l^{'+}}{l^{'-}}\g\g + E\no$~,
\item $e^+e^- \longrightarrow \tilde \chi^0_2 \tilde \chi^0_2
\longrightarrow l^+l^-jj~\g\g + E\no$~,
\item $e^+e^- \longrightarrow \tilde \chi^0_2 \tilde \chi^0_2
\longrightarrow jj~\g\g + E\no$~.
\end{enumerate}
The intermediate states for the process (1) can be either $\tilde
\chi^0_2 \longrightarrow \tilde l^- l^+(\tilde l^{'-} l^{'+})$, 
$\tilde l^- (\tilde l^{'-})\longrightarrow \tilde \chi^0_1 l^-
(l^{'-})$, $\tilde \chi^0_1 \longrightarrow \g \tilde G$ or 
$\tilde \chi^0_2 \longrightarrow \tilde \chi^0_1 Z$, $Z 
\longrightarrow l^+ l^-(l^{'+}l^{'-})$, $\tilde \chi^0_1 
\longrightarrow \g \tilde G$. For process (2) they are $\tilde
\chi^0_2 \longrightarrow \tilde l^- l^+$, $\tilde l^- \longrightarrow
\tilde \chi^0_1 l^-$ and $\tilde \chi^0_2 \longrightarrow \tilde
\chi^0_1 Z$, $Z \longrightarrow q \overline q$. In case of process 
(3) they can be $\tilde \chi^0_2 \longrightarrow \tilde \chi^0_1 Z$, 
$Z \longrightarrow \n \overline \n$ or $q \overline q$. 
The cross sections for the three processes have been plotted 
as functions of $\L$ in Figs. 4 and 5, for two different combinations 
of other parameters. Here we have performed a parton level Monte 
Carlo calculation, setting the criterion for jet merging as 
$\Delta R \geq 0.7,$ where $(\Delta R)^2 = (\Delta \eta)^2 + 
(\Delta \phi)^2$, $\eta$ and $\phi$ being pseudorapidity and  
azimuthal angle, respectively. We have employed a cut of 15 GeV on 
jet energy and an angular cut ${10}^0 < \theta_j < {170}^0$.
For the observation of the two emitted photons, a minimum energy cut of
10 GeV and an angular cut ${10}^0 < \theta_{\g} < {170}^0$ have been
used. A similar angular cut on the emitted leptons has been applied 
(the last-mentioned cut does not affect the cross sections in a
significant manner). The standard model backgrounds for all the 
signals are rather small \cite {ruckl}. 

In Fig. 4, we see that channel (3) dominates over the others for
higher values of $\L$ due to the production of real Z. Conversely, 
channel (1) is more important for lower values of $\L$, when only real  
(right)sleptons are produced. If real Z's are produced at all in such
cases, they are kinematically suppressed.  In Fig. 5, the 
cross sections are quite high $(\sim 0.1 {\rm pb})$ for channel (1) 
compared to the other channels. In this case, $\tilde \chi^0_2$ is 
mainly a gaugino with a high production rate, decaying into real 
sleptons.

The possible sources of confusion with MSSM are 
$\tilde \chi^0_3 \tilde \chi^0_3$ or $\tilde \chi^0_2 \tilde 
\chi^0_3$ pairs. The subsequent decay of $\tilde \chi^0_3$ via the 
channel $\tilde \chi^0_3 \longrightarrow \tilde \chi^0_2 Z$ or 
$\tilde \chi^0_3 \longrightarrow f \tilde f, \tilde f 
\longrightarrow \tilde \chi^0_2 f$ and then the radiative decay
of $\tilde \chi^0_2$ into a photon and a $\tilde \chi^0_1$ can give rise
to the signals mentioned above. These backgrounds are not very serious
due to the following qualitative argument:
the decay $\tilde \chi^0_3 \longrightarrow \tilde \chi^0_2 Z$ 
will be suppressed if $\tilde \chi^0_3$ and $\tilde \chi^0_2$ are not
both dominated by the same Higgsino component. Also, there will be a
kinematical suppression for this decay. Again, if $\tilde \chi^0_3$ is
dominated by the Higgsino component then the production of $\tilde
\chi^0_3$ pair will be very low. The same line of reasoning goes for the
$\tilde \chi^0_3$$\tilde \chi^0_2$ pair production and their subsequent
decays. Also, the cascade decay of $\tilde \chi^0_3$ via sfermion
channel is unlikely to produce signals of appreciable magnitudes, unless
all of the three lightest neutralinos have sizable gaugino components.

For example, the $jj~\g\g + E\no$ signal can come from the following 
process:
$$e^+e^- \longrightarrow \tilde \chi^0_2 \tilde \chi^0_3
\longrightarrow jj~\g\g + E\no~,$$
where, the intermediate states are $\tilde \chi^0_3 \longrightarrow
\chi^0_2 Z$, $Z \longrightarrow q \overline q$, and $\tilde \chi^0_2 
\longrightarrow \tilde \chi^0_1 \g$. The cross section for this signal 
can be at most 0.6-0.7 fb using the same set of cuts which were earlier
used in identifying the GMSB signal. The cross-sections for the other
processes are always smaller than this one. Thus we see that the signals
coming from the pair-production of $\tilde \chi^0_2$ in GMSB scenario is
always large compared to its MSSM counterpart. This definitely happens
for $\L \leq$ 85 TeV, where signals with rates on the order of 1 fb
and much above are predicted.

\section{Summary and Conclusions}

We have considered the effect of the dominant radiative decay of second
lightest neutralino ($\tilde \chi^0_2$) in MSSM, assuming partial
relaxation of gaugino mass unification at a high-energy scale. It has
been shown that this decay may give rise to signals of the type $\g\g +
E\no$ in an $e^+e^-$ collider with $\sqrt s$ = 500 GeV. This in turn can
fake the GMSB scenario where the same signal is supposed to come from 
the decay of the lightest neutralino ($\tilde \chi^0_1$) into a photon 
and a gravitino. Our analysis shows that for some region in the MSSM 
parameter space, this signal is comparable to that coming from GMSB. 
This is particularly true when the mass of $\tilde \chi^0_1$ in the GMSB 
scenario is $\sim 200~GeV$ or more, which marks the region beyond which 
$\tilde \chi^0_1$-pair production is kinematically more and more
suppressed.

The above observations can be generalized to the rather interesting
conclusion that the signals associated with the n{\it th} lightest 
neutralino in a GMSB scenario have always got the chance of being 
faked by the (n+1){\it th} lightest neutralino of MSSM. 
Careful analysis is therefore required to distinguish the two 
possibilities.

In distinguishing between the two scenarios, we see that the idea of
polarization asymmetry is an useful tool. This is actually a measure of
sensitivity of the signal to electron polarization. This idea is 
helpful in separating the two SUSY breaking schemes over a wide region 
of the MSSM parameter space. Also, one can use the combined energy 
distributions of the emitted photons for unpolarized electron beam for
a clear distinction between the two scenarios, particularly when the 
polarization asymmetry is of similar nature for both of them. 

In addition we have also studied the signals coming from the
next-to-lightest neutralino pairs in GMSB. The signals are considerably
free from standard model backgrounds, and contributions from MSSM are
generally suppressed. We conclude that such signals can also be useful
for the confirmation of GMSB in the parameter range upto $\L \approx$
85 TeV.
\vskip .6cm

\noindent {\large \bf Acknowledgments}

\noindent We thank Anirban Kundu for useful discussions. SR wishes to
acknowledge the hospitality of the theory group, Saha Institute of
Nuclear Physics, Calcutta, where part of this work was done.

\newpage

\begin{center}
\begin{tabular}{|c|c|c|c|c|c|c} \hline\hline
{$M_1$} & {$M_2$} & $\s^{\g\g}_{tot}$ & $\s^{\g\g}_L$ &   
$\s^{\g\g}_R$ & $A^{\g\g}$ \\
(GeV) & (GeV) & (fb) & (fb) & (fb) & \\ \hline
70 & 60 & 3.10 & 2.68 & 3.53 & -0.13 \\
80 & 60 & 2.25 & 1.12 & 3.38 & -0.50 \\
90 & 60 & 1.76 & 0.72 & 2.80 & -0.59 \\
90 & 120 & 3.20 & 2.02 & 4.38 & -0.36 \\
100 & 90 & 74.64 & 89.10 & 60.18 & 0.19 \\
100 & 100 & 131.9 & 131.87 & 131.99 & -0.0004 \\
100 & 130 & 8.85 & 5.00 & 12.70 & -0.43 \\
110 & 100 & 68.75 & 85.00 & 52.50 & 0.23 \\
110 & 120 & 74.90 & 60.00 & 89.75 & -0.19 \\
120 & 80 & 43.80 & 80.25 & 7.33 & 0.83 \\ 
120 & 90 & 65.90 & 112.75 & 19.00 & 0.71 \\
140 & 70 & 21.49 & 42.75 & 0.23 & 0.99 \\
140 & 90 & 10.00 & 19.40 & 0.58 & 0.94 \\
150 & 70 & 21.92 & 43.75 & 0.09 & 0.99 \\
150 & 90 & 9.36 & 18.50 & 0.23 & 0.97 \\ \hline
\end{tabular}		      	  	 
\end{center}

\vskip 2 true cm
\centerline{\bf Table 1}

\noindent Sample values of the cross-sections for $e^+e^- 
\longrightarrow \g\g + E\no$, and the asymmetry
parameter in MSSM, for $\mu = -M_Z$, ${\rm tan}\beta$ = 
1.2, $m_{\tilde e}$ = 120 GeV, $m_{\tilde q}$ = 300 GeV.
$\sqrt s$ = 500 GeV.
\newpage
\begin{center}
\begin{tabular}{|c|c|c|c|c|c|c} \hline\hline
{$M_1$} & {$M_2$} & $\s^{\g\g}_{tot}$ & $\s^{\g\g}_L$ &   
$\s^{\g\g}_R$ & $A^{\g\g}$ \\
(GeV) & (GeV) & (fb) & (fb) & (fb) & \\ \hline
50 & 50 & 15.10 & 17.40 & 12.80 & 0.15 \\
50 & 60 & 3.14 & 3.72 & 2.55 & 0.18 \\
60 & 60 & 8.50 & 9.80 & 7.20 & 0.15 \\
70 & 50 & 2.73 & 2.80 & 2.65 & 0.02 \\
70 & 70 & 4.10 & 4.70 & 3.50 & 0.14 \\
70 & 80 & 2.49 & 2.95 & 2.03 & 0.18 \\
100 & 110 & 2.39 & 1.94 & 2.83 & -0.18 \\
100 & 120 & 2.47 & 1.62 & 3.32 & -0.34 \\
110 & 90 & 2.41 & 3.60 & 1.21 & 0.49 \\
110 & 100 & 3.34 & 4.13 & 2.55 & 0.23 \\
110 & 120 & 3.67 & 2.91 & 4.43 & -0.20 \\
120 & 110 & 4.01 & 5.06 & 2.97 & 0.26 \\
160 & 60 & 1.58 & 0.81 & 2.35 & -0.48 \\
170 & 60 & 1.74 & 0.89 & 2.59 & -0.48 \\
180 & 60 & 1.91 & 1.01 & 2.81 & -0.47 \\ \hline
\end{tabular}		      	  	 
\end{center}

\vskip 2 true cm
\centerline{\bf Table 2} 

\noindent Sample values of the cross-sections for $e^+e^- 
\longrightarrow \g\g + E\no$, and the asymmetry
parameter in MSSM, for $\mu = -M_Z$, ${\rm tan}\beta$ = 
1.2, $m_{\tilde e}$ = 1000 GeV, $m_{\tilde q}$ = 1000 GeV.
$\sqrt s$ = 500 GeV.
\newpage
\normalbaselineskip = 16 true pt
\normalbaselines
\setlength{\unitlength}{0.240900pt}
\ifx\plotpoint\undefined\newsavebox{\plotpoint}\fi
\sbox{\plotpoint}{\rule[-0.200pt]{0.400pt}{0.400pt}}%
%

\vskip 2 true cm

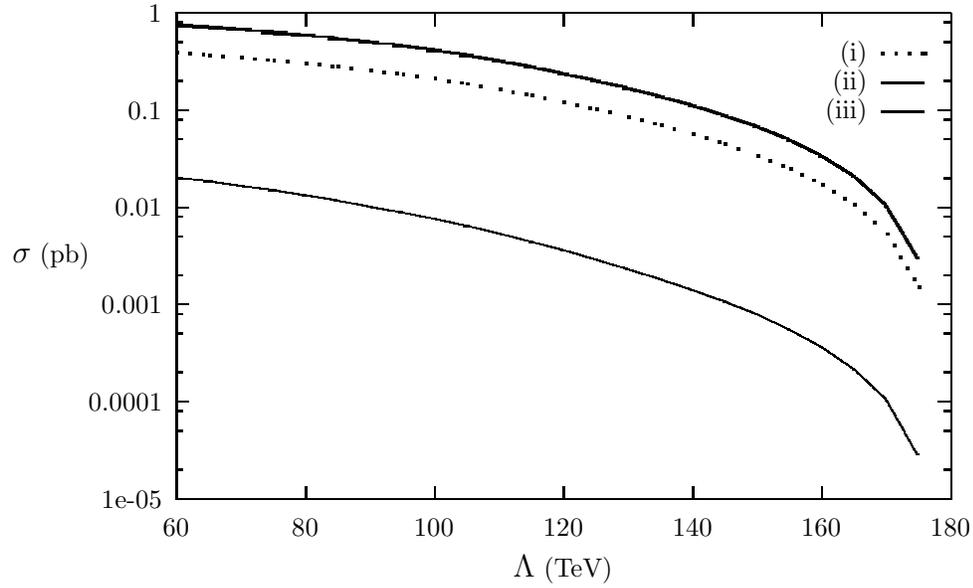
\centerline{\bf Figure 1}

\noindent Cross-section for $e^+e^- \longrightarrow \gamma\gamma + E\no$
~as a function of $\Lambda$ ($\sqrt s$ = 500 GeV). The three lines 
corresponds to (i) unpolarized, (ii) left-polarized, and 
(iii) right-polarized electron beam. Here $\mu$ = 900 GeV, 
${\rm tan}\beta$ = 2, $\frac M \Lambda$ =2. The cuts used here are 
already stated in the text.
\newpage

\setlength{\unitlength}{0.240900pt}
\ifx\plotpoint\undefined\newsavebox{\plotpoint}\fi
\sbox{\plotpoint}{\rule[-0.200pt]{0.400pt}{0.400pt}}%
%

\vskip 2 true cm

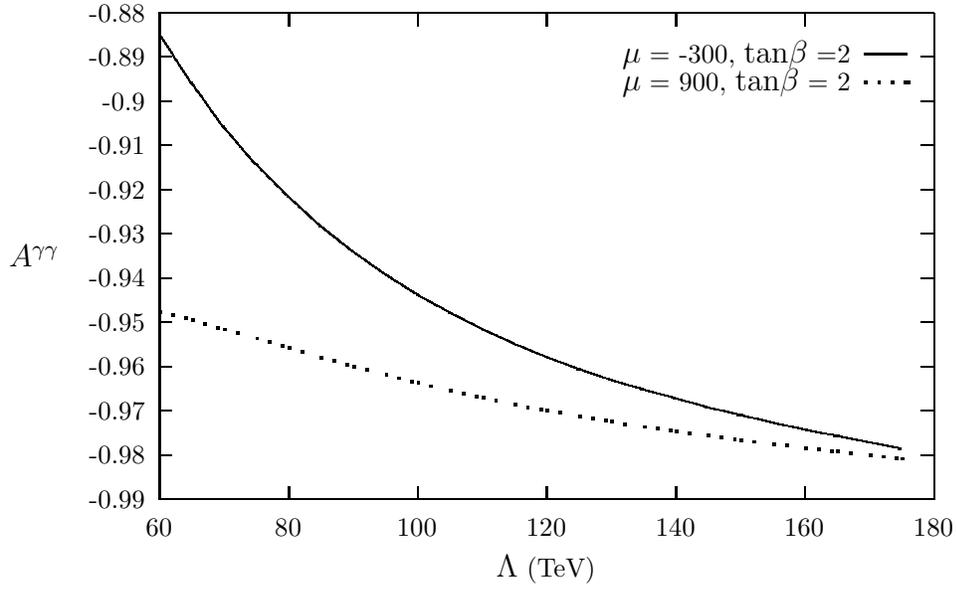
\centerline{\bf Figure 2}

\noindent Asymmetry parameter $A^{\gamma\gamma}$ is plotted as a
function of $\Lambda$ for two different sets of $\mu$ and ${\rm
tan}\beta$. $\frac M \Lambda$ = 2.
\newpage

\setlength{\unitlength}{0.240900pt}
\ifx\plotpoint\undefined\newsavebox{\plotpoint}\fi
\sbox{\plotpoint}{\rule[-0.200pt]{0.400pt}{0.400pt}}%
%

\vskip 2 true cm
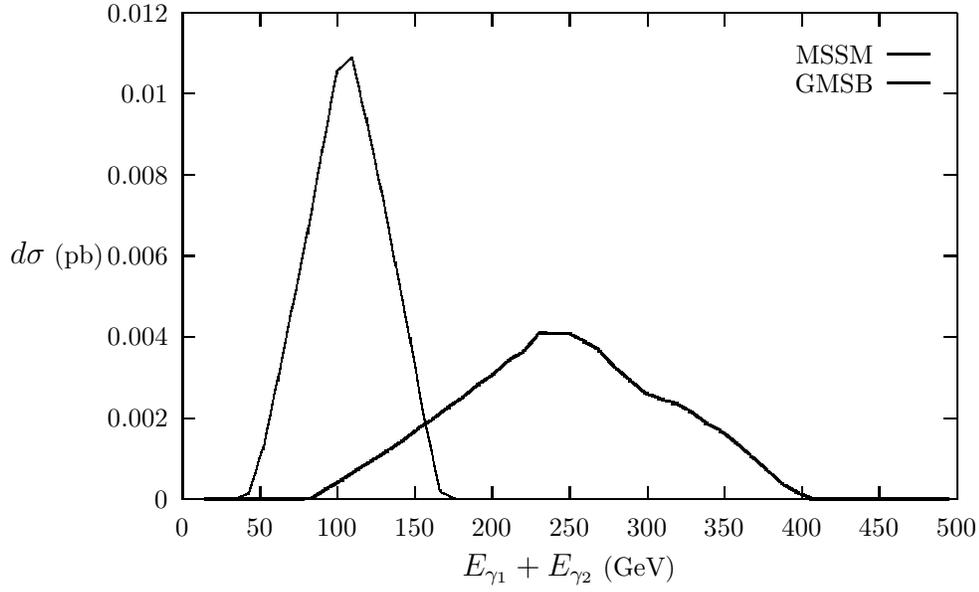
\centerline {\bf Figure 3}

\noindent Total energy distributions of the emitted photons with the 
cuts as stated in the text. For GMSB, $\mu$ = 900 GeV, 
${\rm tan}\beta$ = 2, 
M = 270000 GeV and $\frac M \Lambda$ = 2. For MSSM, $\mu$ = -$M_Z$,
${\rm tan}\beta$ = 1.2, $m_{\tilde e}$ = 120 GeV, $m_{\tilde q}$ = 
300 GeV, $M_2$ = 120 GeV, and $M_1$ = 110 GeV.
\newpage

\setlength{\unitlength}{0.240900pt}
\ifx\plotpoint\undefined\newsavebox{\plotpoint}\fi
\sbox{\plotpoint}{\rule[-0.200pt]{0.400pt}{0.400pt}}%
%

\vskip 2 true cm

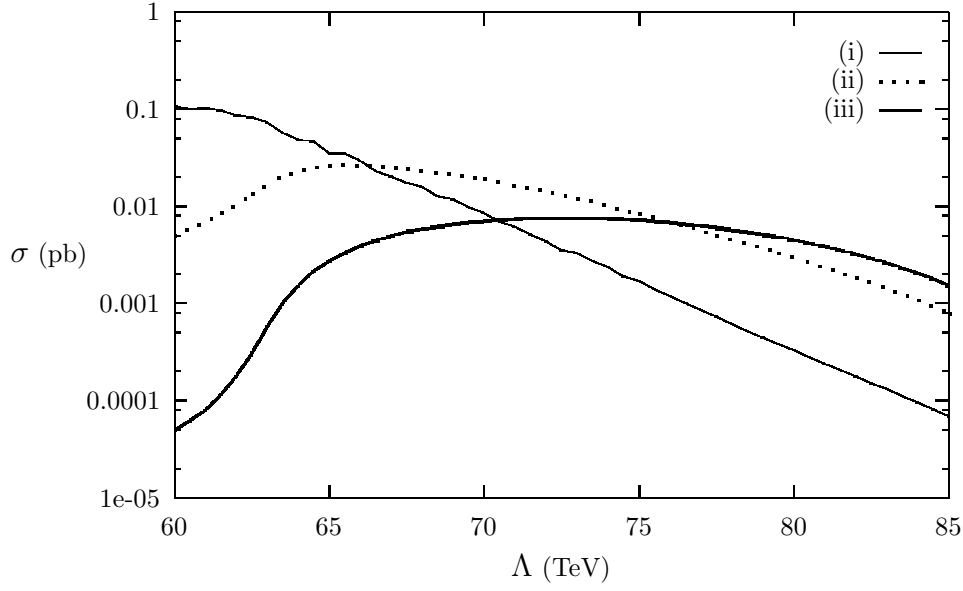
\centerline{\bf Figure 4}

\noindent Cross-sections for (i) $e^+e^- \longrightarrow
l^+l^-l^{'+}l^{'-}\gamma\gamma + E\no$, (ii) $e^+e^- \longrightarrow 
jj~l^+l^-\gamma\gamma + E\no$, and (iii) $e^+e^- \longrightarrow
jj~\gamma\gamma + E\no$~ as functions of $\Lambda$, 
for $\sqrt s$ = 500 GeV. Here $\mu$ = -300, ${\rm tan}\beta$ = 2, 
$\frac M \Lambda$ = 2.
\newpage

\setlength{\unitlength}{0.240900pt}
\ifx\plotpoint\undefined\newsavebox{\plotpoint}\fi
\sbox{\plotpoint}{\rule[-0.200pt]{0.400pt}{0.400pt}}%
%

\vskip 2 true cm

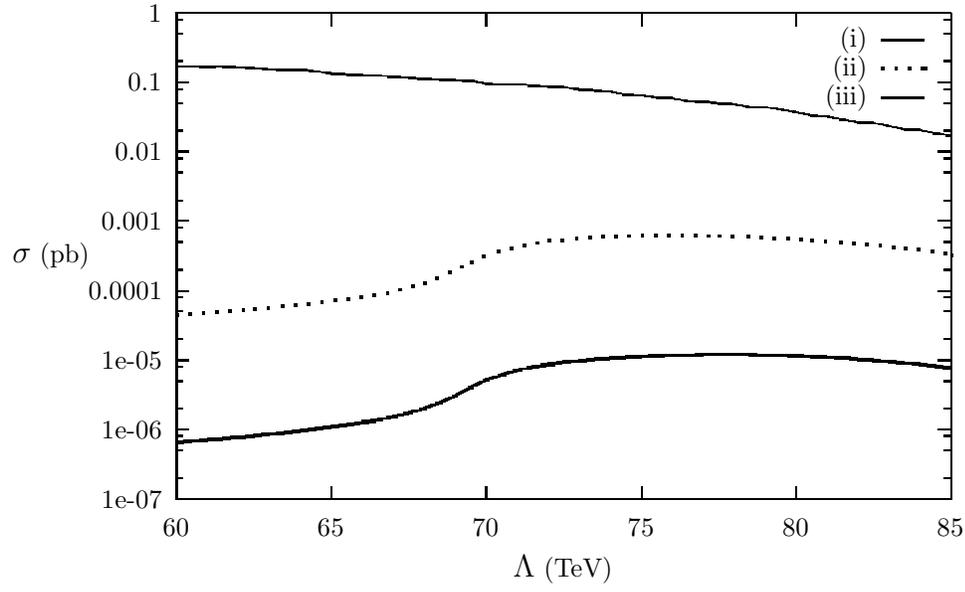
\centerline{\bf Figure 5}

\noindent Cross-sections for (i) $e^+e^- \longrightarrow
l^+l^-l^{'+}l^{'-}\gamma\gamma + E\no$,
(ii) $e^+e^- \longrightarrow jj~l^+l^-\gamma\gamma + E\no$, and 
(iii) $e^+e^- \longrightarrow jj~\gamma\gamma + E\no$~
as functions of $\Lambda$, for $\sqrt s$ = 500 GeV. 
Here $\mu$ = 900, ${\rm tan}\beta$ = 2, $\frac M \Lambda$ = 2.

\end{document}